\documentclass[prb,preprint,groupedaddress, showkeys,showpacs]{revtex4}

\usepackage{graphicx}
\begin{document}


\title{Directed transport driven by L\'{e}vy flights coexisting with subdiffusion}


\author{Bao-quan  Ai$^{1}$}\email[Email: ]{aibq@hotmail.com}
 \author{Ya-feng He$^{2}$}

\affiliation{$^{1}$Laboratory of Quantum Information Technology,
ICMP and
 SPTE, South China Normal University, 510006 Guangzhou, China.\\
 $^{2}$College of Physics Science and
Technology, Hebei University, 071002 Baoding, China}


\date{\today}
\begin{abstract}
\baselineskip 0.4 in
 \indent Transport of the Brownian particles
driven by L\'{e}vy flights coexisting with subdiffusion  in
asymmetric periodic potentials is investigated in the absence of any
external driving forces. Using the Langevin-type dynamics with
subordination techniques, we obtain the group velocity which can
measure the transport. It is found that the group velocity increases
monotonically with the subdiffusive index and there exists an
optimal value of the L\'{e}vy index at which the group velocity
takes its maximal value. There is a threshold value of the
subdiffusive index below which the ratchet effects will disappear.
The nonthermal character of the L\'{e}vy flights and the asymmetry
of the potential are necessary to obtain the directed transport.
Some peculiar phenomena induced by the competition between L\'{e}vy
flights and subdiffusion are also observed. The pseudo-normal
diffusion will appear on the level of the median.
\end{abstract}

\pacs{ 05. 40. Fb, 05. 10. Gg, 05. 40. -a}
\keywords{Ratchet, L\'{e}vy flights, subdiffusion}



\maketitle

\baselineskip 0.4 in
\section {Introduction}
\indent Directed Brownian motion induced by zero-mean
non-equilibrium fluctuations in the absence of macroscopic forces
and potential gradients is presently under intense investigation
\cite{a1}. This comes from the desire to understand unidirectional
transport in biological systems\cite{a2}, as well as their potential
technological applications ranging from classical non-equilibrium
models\cite{a3} to quantum systems\cite{a4}. A ratchet system is
generally defined as a system that is able to transport particles in
a periodic structure with nonzero macroscopic velocity in the
absence of macroscopic force on average. Broadly speaking, ratchet
devices fall into three categories depending on how the applied
perturbation couples to the substrate asymmetry:  rocking
ratchets\cite{a5}, flashing ratchets\cite{a6}, and correlation
ratchets \cite{a7}. Additionally, entropic ratchets, in which
Brownian particles move in a confined structure, instead of a
potential, were also extensively investigated \cite{a8}. These
studies on directed transport of the Brownian  particles focused on
the normal diffusion.\\
\indent However, anomalous diffusion has attracted growing
attention, being observed in various fields of physics and related
sciences\cite{a9}. Description of physical models in terms of
L\'{e}vy flights and subdiffusion becomes more and more popular
\cite{a10,a11,a12,a13,a14,a15,a16,a17,a18,a19,a20,a21,a22,a23}. In
the complex systems the distinct class of subdiffusion processes was
reported in condensed phases\cite{a10}, ecology\cite{a11}, and
biology\cite{a12}. Superdiffusion driven by Levy flights is actually
observed in various real systems and is used to model a variety of
processes such as bulk mediated surface diffusion \cite{a13},
exciton and charge transport in polymers under conformational motion
\cite{a14}, transport in micelle systems or heterogeneous rocks
\cite{a15},two-dimensional rotating flow \cite{a16}, and many
others\cite{a9}.  Goychuk and coworkers \cite{a17} studied the
subdiffusive transport in tilted periodic potentials and established
a universal scaling relation for diffusive transport. Dybiec and
coworkers \cite{a18} studied the minimal setup for a L\'{e}vy
ratchet and found that due to the nonthermal character of the
L\'{e}vy noise, the net current can be obtained even in the absence
of whatever additional time-dependent forces. Del-Castillo-Negrete
and coworkers \cite{a19} also found the similar results in constant
force-driven L\'{e}vy ratchet. Rosa and Beims \cite{a20} studied the
optimal transport and its relation to superdiffusive transport and
L\'{e}vy walks for Brownian particles in ratchet potential in the
presence of modulated environment and external oscillating forces.
We also studied the transport of Brownian particles in the presence
of ac-driving forces and L\'{e}vy flights and multiple current
reversals were observed\cite{a21}

\indent Recently, much attention has been devoted to the competition
between subdiffusion and L\'{e}vy flights. The competition is
conveniently described by the fractional Fokker-Planck equation with
temporal and spatial fractional derivatives \cite{a9}. It is very
difficult to see this competition in the framework of the fractional
Fokker-Planck dynamics. Magdziarz and coworkers \cite{a22} proposed
a equivalent approach based on the subordinated Langevin method to
visualize the competition on the level of sample paths as well as on
the level of probability density functions. Based on this approach
Dybiec and coworkers \cite{a23} found that due to the competition
between L\'{e}vy flights and subdiffusion, the standard measure used
to discriminate between anomalous and normal behavior cannot be
applied straightforwardly. Koren and coworkers\cite{a24} have
investigated the first passage times in one-dimensional system
displaying a competition between subdiffusion and L\'{e}vy flights
and found some peculiar phenomena.

\indent What will happen when the particles move in a ratchet
potential subjected to subdiffusion and L\'{e}vy flights? In order
to answer this question we use the subordinated Langevin method
proposed by the Magdziarz and coworkers \cite{a22} to investigate
this competition in a minimal L\'{e}vy ratchet without any external
driving forces. We emphasize on visualizing the competition on the
level of the group velocity and diffusion and finding how this
competition affects the directed transport.

\section{Model and Methods}

\indent We consider the transport of the Brownian particles driven
by L\'{e}vy flights and subdiffusion in the absence of whatever
additional time-dependent forces. The competition between L\'{e}vy
flights and subdiffusion in a ratchet potential $V(x)$ can be
described by the fractional Fokker-Planck equation with temporal and
spatial fractional derivatives \cite{a9, a22}
\begin{equation}\label{}
    \frac{\partial p(x,t)}{\partial t}={_0}D_{t}^{1-\alpha}\left[\frac{\partial }{\partial
    x}\frac{V^{'}(x)}{\eta}+D\frac{\partial^{\mu}}{\partial |x|^{\mu}}\right] p(x,t),
\end{equation}
where $p(x,t)$ is the probability density for particles at position
$x$ and time $t$. The prime stands for the derivative with respect
to the space variable $x$. $D$ is the anomalous diffusion
coefficient which describes the noise intensity in the subordinated
process. $\eta$ denotes the generalized friction constant. Here
${_0}D_{t}^{1-\alpha}$ is the fractional of the Riemann-Liouville
operator ($0<\alpha\leq 1$) defined through \cite{a9, a22}
\begin{equation}\label{}
{_0}D_{t}^{1-\alpha}g(t)=\frac{1}{\Gamma(\alpha)}\frac{d}{dt}\int^{t}_{0}(t-s)^{\alpha-1}g(s)ds,
\end{equation}
where $\Gamma(.)$ is the Gamma function. From the definition, it
becomes apparent that subdiffusion corresponds to a slowly decaying
memory integral in the dynamical equation for $p(x,t)$. The operator
$\frac{\partial^{\mu}}{\partial |x|^{\mu}}$, $0<\mu\leq 2$, stands
for the Riesz fractional derivative \cite{a9, a22} with the Fourier
transform $\mathcal{F}\{\frac{\partial^{\mu}}{\partial |x|^{\mu}}
f(x)\}=-|k|^{\mu}\tilde{f}(k)$.

\indent The occurrences of the operator ${_0}D_{t}^{1-\alpha}$ and
$\frac{\partial^{\mu}}{\partial |x|^{\mu}}$ are induced by the
heavy-tailed waiting times between successive jumps  and the
heavy-tailed distributions of the jumps, respectively, in the
underlying continuous-time random walk scheme. The case of
$\alpha=1$, $\mu=2$ corresponds to the standard Fokker-Planck
equation.  $V(x)$ is an asymmetric periodic potential
\begin{equation}\label{}
    V(x)=\frac{V_{0}}{2\pi}\left[\sin(2\pi x)+\frac{\Delta}{4}\sin(4\pi
    x)\right],
\end{equation}
where $V_{0}$ and $\Delta$  are the amplitude and the asymmetric
parameter of the potential, respectively.

\indent Because it is very difficult to solve Eq. (1) analytically
and numerically, we used the subordinated Langevin method proposed
by Magdziarz and coworkers \cite{a22} to investigate the transport.
In their method, the solution $p(x,t)$ of Eq. (1) is equal to the
probability density function of the subordinated process
\begin{equation}\label{}
    Y(t)=X(S_{t}),
\end{equation}
where the parent process $X(\tau)$ is defined as the solution the
stochastic differential equation
\begin{equation}\label{}
    dX(\tau)=-\frac{V^{'}(X(\tau))}{\eta}d\tau+D^{1/\mu}dL_{\mu}(\tau),
\end{equation}
where $L_{\mu}(\tau)$ is the symmetric $\mu$-stable L\'{e}vy motion
with the Fourier transform
$\mathcal{F}\{L_{\mu}(\tau)\}=e^{-\tau|k|^{\mu}}$. Employing the
Euler scheme to Eq. (5), one can obtain
\begin{equation}\label{}
    X(\tau_{0})=0,
\end{equation}
\begin{equation}\label{}
    X(\tau_{i})=X(\tau_{i-1})-\frac{V^{'}(X(\tau_{i-1}))}{\eta}\Delta
    \tau +(D\Delta\tau)^{1/\mu}\xi_{i},
\end{equation}
where $i=1,2,3...$ and $\xi_{i}$ are the random variables with
standard symmetric $\mu$-stable distribution. The procedure of
generating realizations $\xi_{i}$ is the following \cite{a22,b1}
\begin{equation}\label{}
\xi_{i}=\frac{\sin(\mu V)}{(\cos
V)^{1/\mu}}\left[\frac{\cos([1-\mu]V)}{W}\right]^{\frac{1-\mu}{\mu}},
\end{equation}
where  the random variable $V$ is uniformly distributed on$( -\pi/2,
\pi/2)$, $W$ has exponential distribution with mean one.

The inverse-time $\alpha$-stable subordinator $S_{t}$, which is
assumed to be independent of $X(\tau)$,  is defined as
\begin{equation}\label{}
    S_{t}=\inf\{\tau: U(\tau)>t\},
\end{equation}
where $U(\tau)$ is the strictly increasing $\alpha$-stable L\'{e}vy
motion with Laplace transform $\mathcal{L}\{U(\tau)\}=e^{-\tau
k^{\alpha}}$.

\indent Using the standard method of summing increments of the
process $U(\tau)$ one can get
\begin{equation}\label{}
    U(\tau_{0})=0,
\end{equation}
\begin{equation}\label{}
    U(\tau_{j})=U(\tau_{j-1})+\Delta\tau^{1/\alpha}\zeta_{j},
\end{equation}
where $j=1, 2, 3...$ and $\zeta_{j}$ are the skewed positive
$\alpha$-stable random variables \cite{a22,b1}. The method to
generate the random variables is
\begin{equation}\label{}
    \zeta_{j}=\frac{\sin(\alpha(V+\frac{\pi}{2}))}{[\cos(V)]^{\frac{1}{\alpha}}}\left[\frac{\cos(V-\alpha(V+\frac{\pi}{2}))}{W}\right]^\frac{1-\alpha}{\alpha},
\end{equation}
where $V$ and $W$ have the same definitions as that in Eq. (8). From
the above procedures, one can obtain the subordinated process $Y(t)$
and its probability distribution function is equal to the solution
of Eq. (1). For more detailed information on the algorithm, please
see the Ref. (22).

\indent In the classical ratchets, one can use the average velocity
and effective diffusion coefficient to describe the transport.
However, for the noise with distribution of a L\'{e}vy-stable law,
the mean of the noise  and the second moment may do not exist.  As a
consequence, the classical stochastic theory (average velocity and
effective diffusion coefficient), which is based on the ordinary
central limit theorem, is no longer valid. To overcome this problem,
Dybiec and coworkers \cite{a18} recently proposed a different
approach to the L\'{e}vy ratchet problem based on the quantile line
analysis for $0<\mu<2$.

\indent Quantile line is a very useful tool for investigation of the
overall motion of the probability density of finding a particle in
the vicinity of $Y(t)$ \cite{a18,a23}.  A median line for a
stochastic process $Y(t)$ is a function of $q_{0.5}(t)$ given by the
relationship $Pr(Y(t)\leq q_{0.5}(t))=0.5$. Therefore, one can use
the derivative of the median to define the group velocity of the
particle packet\cite{a18},
\begin{equation}\label{}
    V_{g}=\frac{d q_{0.5}(t)}{dt},
\end{equation}
and this definition is valid even for the case of lacking average
velocity.

\section {Numerical results and discussion}
\indent In order to investigate the competition between L\'{e}vy
flights and subdiffusion in a ratchet potential, we carried out
extensively numerical simulations based on the subordinated Langevin
method \cite{a22}. For simplicity we set $\eta=1.0$ and $1<\mu\leq
2$ throughout the work.  In our simulations, we have considered more
than $10^{5}$ realizations to obtain the accurate median. In order
to provide the requested accuracy of the system the dynamics time
step was chosen to be smaller than $10^{-3}$. We have checked that
these are sufficient for the system to obtain consistent results.


\indent Firstly, we will investigate the diffusive properties of the
Brownian particles.  Usually, the types of the diffusion processes
are determined by the spread of the distance traveled by a random
walker. The diffusion is characterized through the power law form of
the mean-square displacement $\langle x^{2}(t)\rangle\propto
t^{\delta}$. According to the value of the index $\delta$, one can
distinguish subdiffusion ($0<\delta<1$), normal diffusion
($\delta=1$) and superdiffusion ($\delta>1$). Here, we use the
median of square displacement $M(x^{2})$, instead of mean-square
displacement, to characterize the diffusion. Fig. 1 (a) shows the
time dependence of $M(x^{2})/t$ for different combinations of $\mu$
and $\alpha$ without any external potential. It is found that the
linear time dependence of the median of square displacement,
$M(x^{2})\propto t$, will occur for the case of
$\frac{2\alpha}{\mu}$=1, which indicates the normal diffusion.
However, this is not true, for example $\mu=1.8$ and $\alpha=0.9$,
the process is still non-Markov and non-Gaussian. This pseudo-normal
diffusion is due to the competition between L\'{e}vy flights and
subdiffusion. Dybiec and coworkers \cite{a23} have presented
discussions in detail on this paradoxical diffusion. Fig. 1 (b)
presents the time dependence of $M(x^{2})/t$ in the presence of a
ratchet potential. Interestingly, the pseudo-normal diffusion for
$\mu=1.8$ and $\alpha=0.9$ with external potentials is not normal.


\indent Next, we will study the rectified mechanism of the L\'{e}vy
ratchets.  Usually, the ratchet mechanism demands three key
ingredients \cite{b2} which are (a) nonlinear periodic potential: it
is necessary since the system will produce a zero mean output from
zero-mean input in a linear system; (b)asymmetry of the potential,
it can violate the symmetry of the response; (c)fluctuating:
L\'{e}vy flights can break thermodynamical equilibrium. In Fig. 2
(a), we studied the time dependence of the median for different
values of the asymmetry parameter $\Delta$ at $\mu=1.5$ and
$\alpha=1.0$. The median is positive for $\Delta>0$, zero at
$\Delta=0$, and negative for $\Delta<0$. Therefore, the asymmetry of
the potential will determine the direction of the transport and no
directed transport occurs in a symmetric potential. Now we will give
the physical interpretation of the directed transport for the case
of $\Delta=1$. Firstly, the particles stay in the minima of the
potential awaiting large noise pulse to be catapulted out. The
particles will be thrown out to the left and the right with the
equal probabilities. In this case, the distance from minima to
maxima is shorter from the right side than that from the left side.
Consequently, most of the particles are thrown out from the right
side, resulting in positive transport. This gives rise to the
overall preferred motion to the right.

 \indent Figure 2 (b) gives the time dependence of the median for
 different combinations of $\mu$ and $\alpha$. We find that L\'{e}vy
 flights are necessary to obtain the directed transport. For
 Gaussian case ($\mu=2.0$), directed transport disappears.
 This is due to the nonthermal character of the L\'{e}vy flights that can break
 thermodynamical equilibrium. From Fig. 2(a) and (b) we can see that
 the asymmetry of the potential and the non-equilibrium character of
 the L\'{e}vy flights are the two necessary conditions for directed
 transport. The direction of the transport is determined by the
 direction of the steeper slope of of the potential and the L\'{e}vy flights
 can break thermaldynamical equilibrium. These two key
ingredients can realize the ratchet effects.

\indent Figure 3 illustrates the dependence of the group velocity
$V_{g}$ on the subdiffusive index $\alpha$ for different values of
 the L\'{e}vy index $\mu$. One can see that group velocity $V_{g}$
 increases monotonically with the subdiffusive index $\alpha$.  For small values of
 $\alpha$, the waiting time between successive jumps is very long
 and  it is not easy for particles to pass across the barrier. Thus, most particles will stay in their original minima of the
 potential and the group velocity becomes very small. Especially, we
 also find that there exists a threshold value of $\alpha$ below
 which no directed transport can be obtained. The
 subdiffusion dominates the transport for small values of $\alpha$
 ($\alpha<0.7$), while the effects of the L\'{e}vy flights become preponderant
 for large values of $\alpha$.

\indent Figure 4 shows the dependence of the group velocity $V_{g}$
on the L\'{e}vy index $\mu$ for different values of $\alpha$. When
$\mu\rightarrow 2.0$, the system is under thermodynamical
equilibrium and no directed transport appears. For small values of
$\mu$, L\'{e}vy flights are longer and the outliers in the L\'{e}vy
noise are larger. In this case, the effects of the asymmetry of the
potential become very small, resulting in small group velocity.
Therefore, there exists a optimal value of $\mu$ at which the group
velocity takes its maxima. This can also be confirmed by Fig. 3. For
very small values of $\alpha$, for example $\alpha=0.5$, the group
velocity is zero for all values of $\mu$ and the transport is
absolutely dominated by subdiffusion.

\indent The group velocity $V_{g}$ as a function of noise intensity
$D$ is shown in Fig. 5 for different combinations of $\mu$ and
$\alpha$. The curve is observed to be bell shaped which shows the
feature of resonance. When $D\rightarrow 0$, the particles cannot
pass across the barrier and there is no directed current. When
$D\rightarrow \infty$ so that the noise is very large, the effect of
the potential disappears and the group velocity tends to zero, also.
There is an optimal value of $D$ at which the group velocity is
maximal. There are two intersections ($D_{c1}$ and $D_{c2}$) between
the line of $\mu=1.9$ and $\alpha=1.0$ and the line of $\mu=1.5$ and
$\alpha=0.9$. For simplicity, we define $V_{g}(\mu, \alpha)$ as the
group velocity for different values of $\mu$ and $\alpha$. When
$D<D_{c1}$, $V_{g}(1.9, 1.0)>V_{g}(1.5, 0.9)$, L\'{e}vy flights
dominates the transport. When $D_{c1}<D<D_{c2}$, $V_{g}(1.9,
1.0)<V_{g}(1.5, 0.9)$, the transport is governed by subdiffusion.
For the case of $D>D_{c2}$, all particles can easily pass across the
barrier and L\'{e}vy flights will mainly contribute to the transport
and $V_{g}(1.9, 1.0)>V_{g}(1.5, 0.9)$

\indent In Fig. 6, we plot the dependence of the group velocity
$V_{g}$ on the amplitude $V_{0}$ of the potential for different
combinations of $\mu$ and $\alpha$.  When $V_{0}\rightarrow 0$, the
effects of the potential disappear and the group velocity tends to
zero. When $V_{0}\rightarrow \infty$, the particles cannot pass
across the barrier and the group velocity goes to zero, also. Thus,
the curve shows a peak.  Remarkably, there is an intersection
between the line of $\mu=1.5$ and $\alpha=0.9$ and the line of
$\mu=1.9$ and $\alpha=1.0$. This is due to the competition between
L\'{e}vy flights and subdiffusion. When $V_{0}<V_{c}$, L\'{e}vy
flights are predominant and $V_{g}(1.5, 0.9)>V_{g}(1.9,1.0)$. When
$V_{0}>V_{c}$, subdiffusion dominates the transport and $V_{g}(1.5,
0.9)<V_{g}(1.9, 1.0)$. In this case, the height of the barrier is
very high and few particles driven by L\'{e}vy flights can pass
across the barrier, the effects of the L\'{e}vy flights will
disappear and subdiffusion will play a major role.

\section{Concluding Remarks}
\indent In this paper, we have investigated the directed transport
of the Brownian particles in a ratchet potential driven by L\'{e}vy
flights coexisting with subdiffusion. We used recently developed
framework of Monte Carlo simulation \cite{a22} which is equal to the
solution of the fractional Fokker-Planck equation. The group
velocity proposed by Dybiec and coworkers \cite{a18} is used to
measure the transport. It is found that the group velocity increases
monotonically with the subdiffusive index, while the group velocity
as a function of the L\'{e}vy index is nonmonotonic. The former is
caused by the increase of the waiting time between successive jumps
and the latter is owing to the interplay between L\'{e}vy flights
and the height of the barriers. There is a threshold value of
$\alpha$ below which the transport is absolutely dominated by
subdiffusion and the directed transport disappears. The dependences
of the group velocity on the noise intensity and the amplitude of
the potential are also investigated. There is an optimal value of
the noise intensity (the amplitude of the potential) at which the
group velocity is maximal. The competition between Levy fights and
subdiffusion in the ratchet potential is observed on the level of
the group velocity as well as the median of square displacement.
 The nonthermal character of the L\'{e}vy flights and the asymmetry of the potential are the
necessary conditions for directed transport when the system is in
the absence of any external driving forces. Because of this
competition, we also found the pseudo-normal diffusion reported by
Dybiec and coworkers \cite{a23}, in which time dependence of the
median of square displacement is linear, $M(x^{2})\propto t$, while
the process is still non-Markov and non-Gaussian.

\indent Anomalous transport is becoming widely recognized in a
variety of the fields. Beyond its intrinsic theoretical interest,
the results we have presented may have wide applications in some
complex systems, such as diffusive transport in plasmas, particles
separation with non-Gaussian diffusion, and ratchet transport in
biology systems that are intrinsically out of equilibrium.

\indent We would like to thank Dr. Magdziarz for enthusiastic help
on numerical algorithm. This work was supported in part by National
Natural Science Foundation of China with Grant Nos. 30600122 and
10947166 and GuangDong Provincial Natural Science Foundation with
Grant No. 06025073. Y. F. He also acknowledges the Research
Foundation of Education Bureau of Hebei Province, China (Grant No.
2009108)

\newpage
\section {Caption list}
\baselineskip 0.4 in

Fig. 1. Time dependence of $M(x^{2})/t$ for different combinations
of $\mu$ and $\alpha$: (a)without external potential at
  $D=0.4$, solid lines present $t^{\frac{2\alpha}{\mu}-1}$ scaling;  (b)with external potential at $D=0.4$, $V_{0}=5.0$, and
  $\Delta=1.0$.\\

Fig. 2. Time dependence of the median: (a) for different values of
the asymmetry parameter
  $\Delta$ at $D=0.4$, $V_{0}=5.0$, $\mu=1.5$, and $\alpha=1.0$,  the inset shows the potential profile;
  (b)for different combinations of $\mu$ and $\alpha$ at $D=0.4$, $V_{0}=5.0$, and
  $\Delta=1.0$. \\

Fig. 3. Group velocity $V_{g}$ versus subdiffusive index $\alpha$
for different values of $\mu$ at $D=0.4$, $V_{0}=5.0$, and
$\Delta=1.0$.\\

Fig. 4. Group velocity $V_{g}$ versus L\'{e}vy index $\mu$ for
different values of $\alpha$ at $D=0.4$, $V_{0}=5.0$, and
$\Delta=1.0$. \\

Fig. 5. Group velocity $V_{g}$ as a function of noise intensity $D$
for different combinations of $\mu$ and $\alpha$ at $V_{0}=5.0$ and
$\Delta=1.0$. \\

Fig. 6. Group velocity $V_{g}$ versus the amplitude $V_{0}$ of the
potential for different combinations of $\mu$ and $\alpha$ at
$D=0.4$ and $\Delta=1.0$.\\

\newpage
\begin{figure}[htbp]
  \begin{center}\includegraphics[width=10cm,height=8cm]{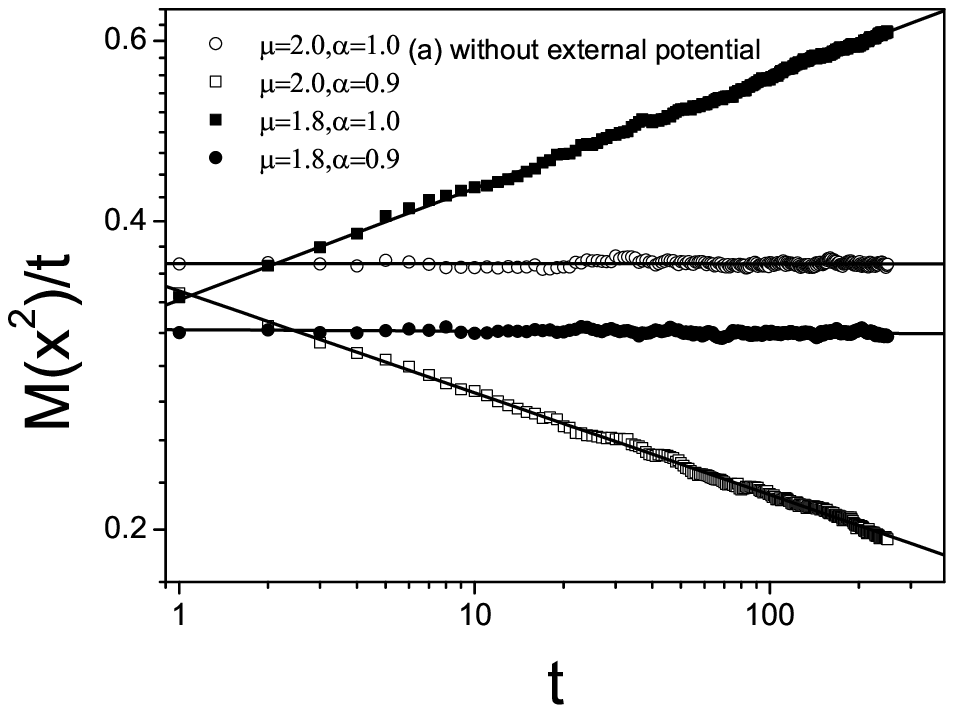}
  \includegraphics[width=10cm,height=8cm]{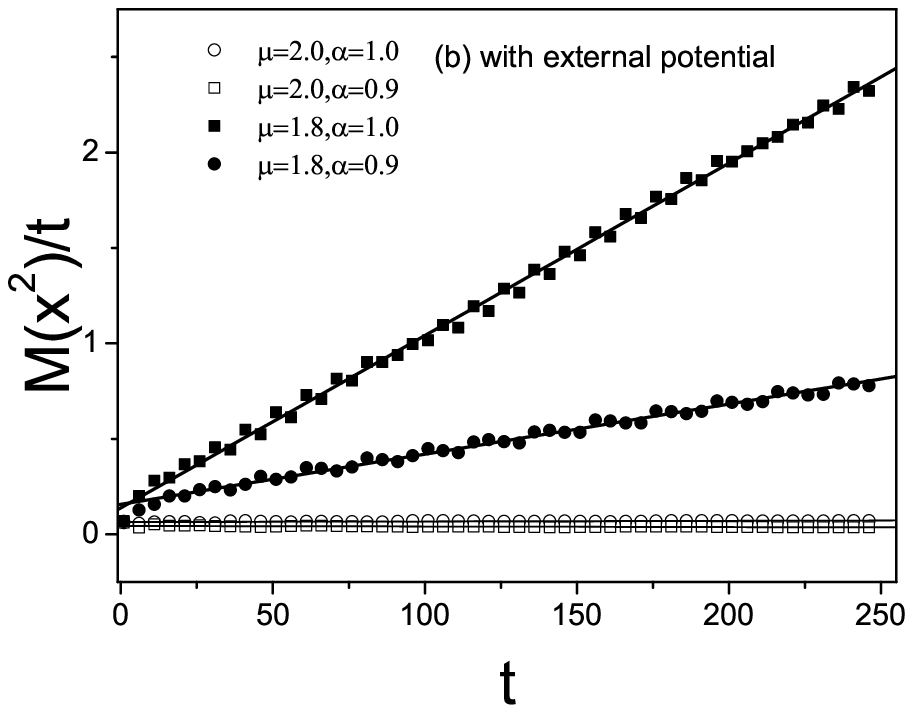}
  \caption{}\label{1}
\end{center}
\end{figure}
\newpage
\begin{figure}[htbp]
  \begin{center}\includegraphics[width=10cm,height=8cm]{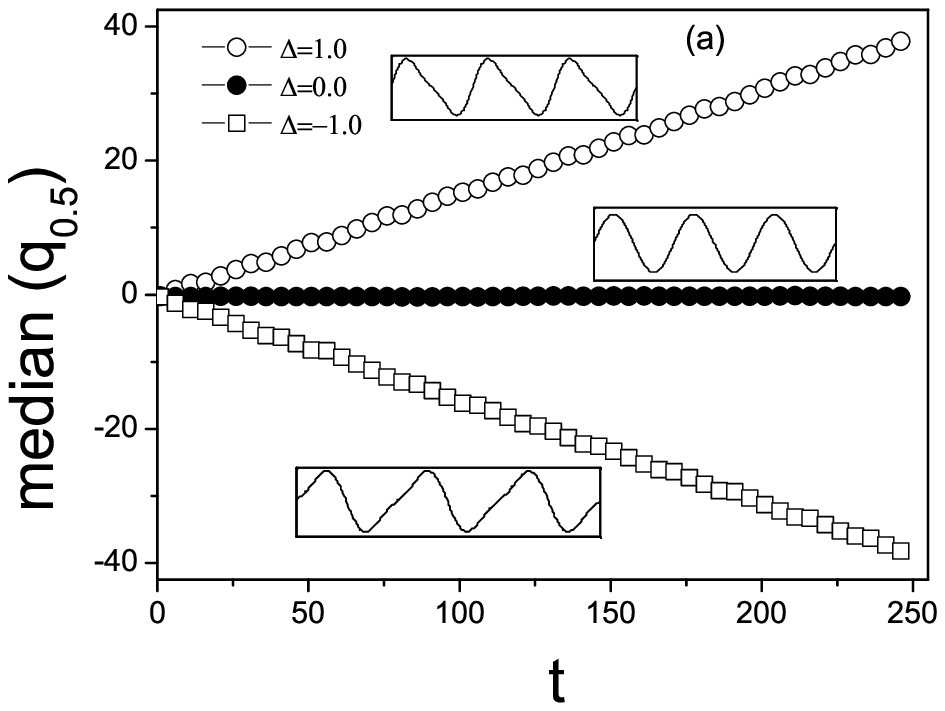}
  \includegraphics[width=10cm,height=8cm]{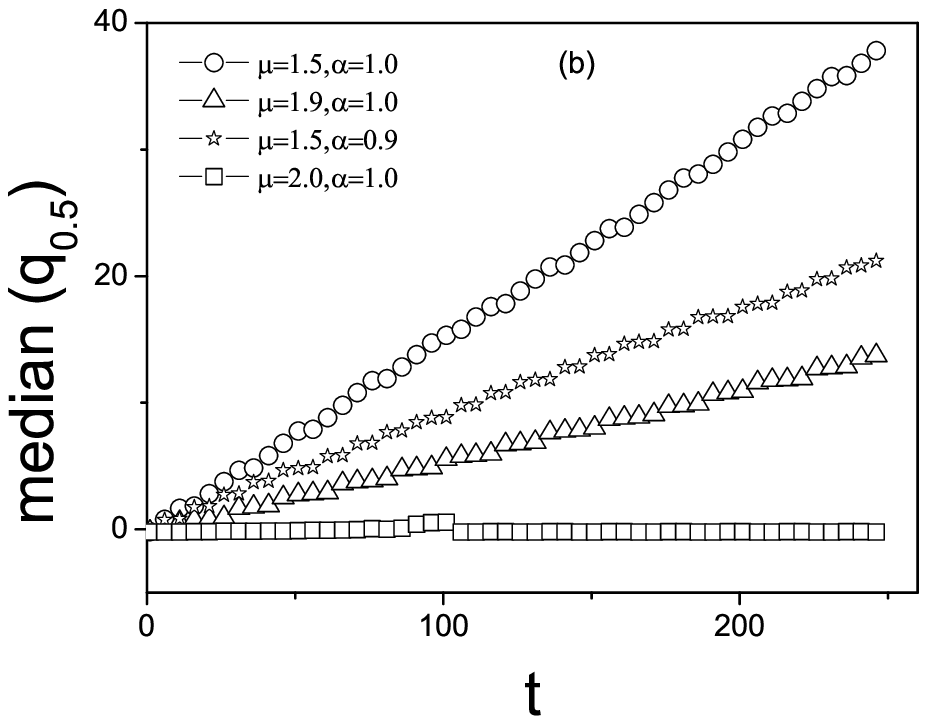}
  \caption{}\label{1}
\end{center}
\end{figure}
\newpage
\begin{figure}[htbp]
  \begin{center}\includegraphics[width=10cm,height=8cm]{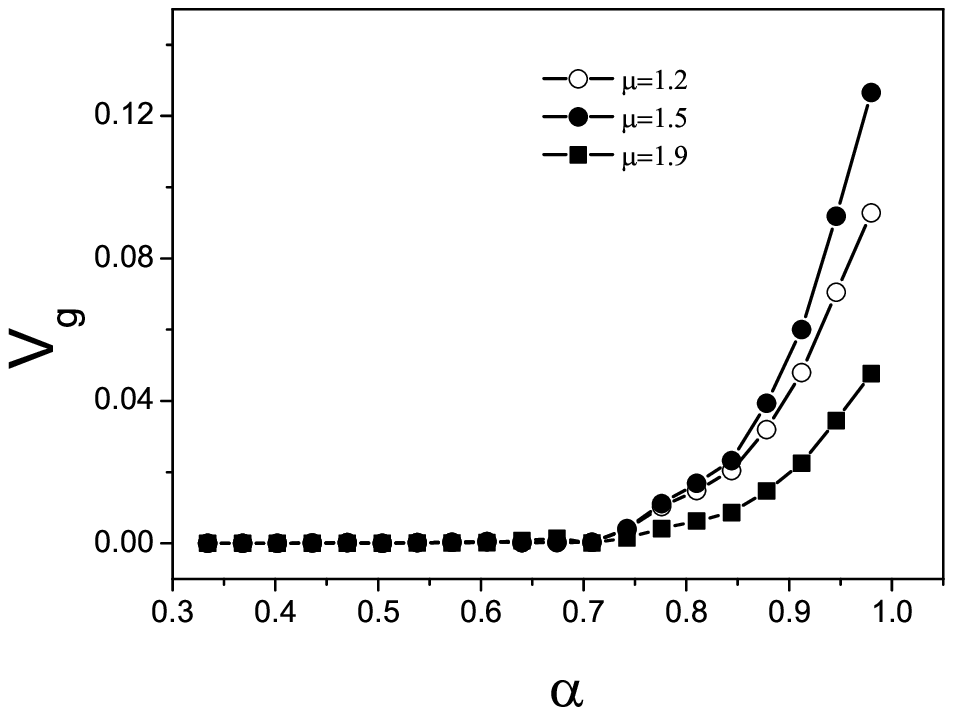}
  \caption{}\label{1}
\end{center}
\end{figure}
\newpage
\begin{figure}[htbp]
  \begin{center}\includegraphics[width=10cm,height=8cm]{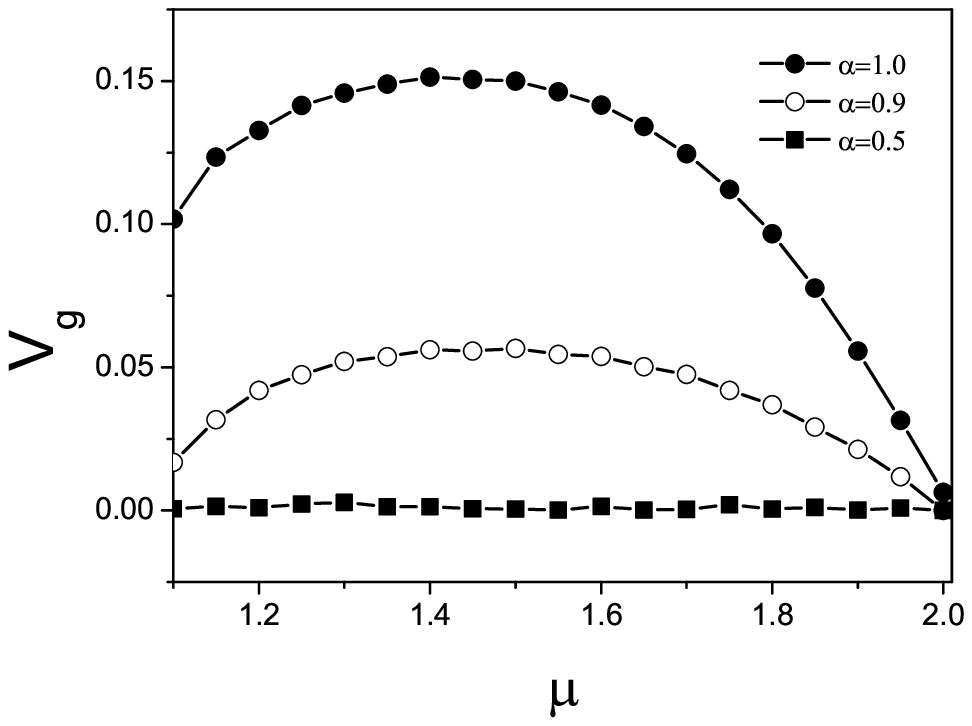}
  \caption{ }\label{1}
\end{center}
\end{figure}
\newpage
\begin{figure}[htbp]
  \begin{center}\includegraphics[width=10cm,height=8cm]{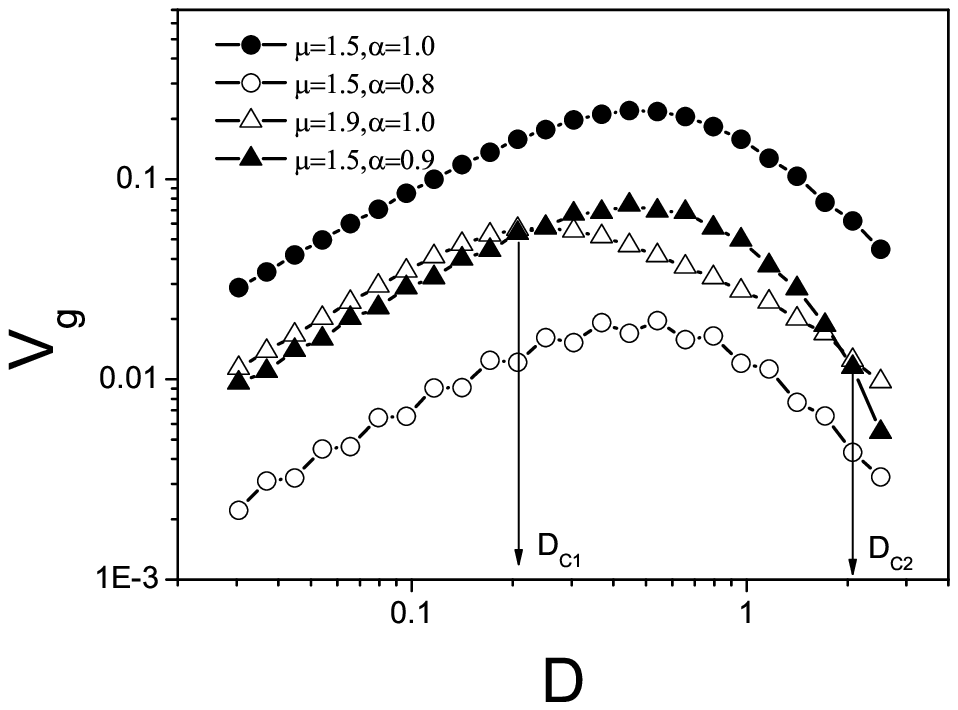}
  \caption{}\label{1}
\end{center}
\end{figure}
\newpage
\begin{figure}[htbp]
  \begin{center}\includegraphics[width=10cm,height=8cm]{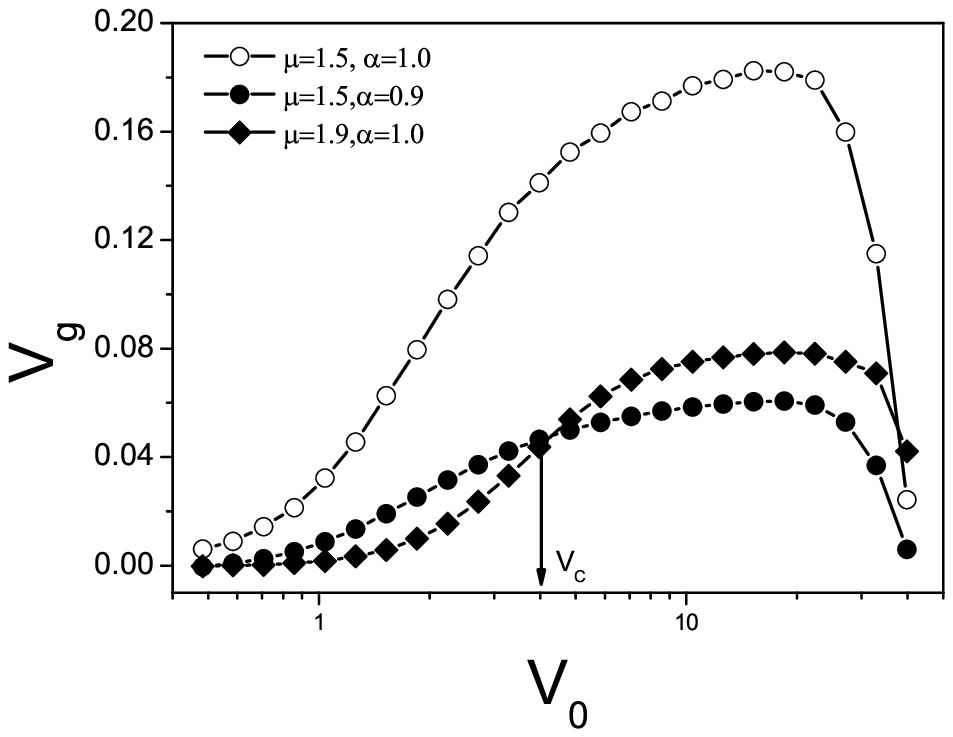}
  \caption{}\label{1}
\end{center}
\end{figure}

\end{document}